\documentclass[12pt,thmsa]{article}

\usepackage{cite}
\usepackage{graphicx}
\usepackage{dcolumn}

\begin{document}

\title{Rational Approximation for Two--Point Boundary Value Problems}
\author{Paolo Amore\thanks{
paolo@ucol.mx} \and Facultad de Ciencias, Universidad de Colima, \and Bernal
D\'iaz del Castillo 340, Colima, Colima, Mexico \and Francisco M.
Fern\'andez \thanks{
Correspondent author} \thanks{
fernande@quimica.unlp.edu.ar} \\
INIFTA (Conicet, UNLP), Divisi\'on Qu\'imica Te\'orica, Diag 113 S/N,\\
Sucursal 4, Casilla de Correo 16, 1900 La Plata, Argentina}
\maketitle

\begin{abstract}
We propose a method for the treatment of two--point boundary value problems
given by nonlinear ordinary differential equations. The approach leads to
sequences of roots of Hankel determinants that converge rapidly towards the
unknown parameter of the problem. We treat several problems of physical
interest: the field equation determining the vortex profile in a
Ginzburg--Landau effective theory, the fixed--point equation for Wilson's
exact renormalization group, a suitably modified Wegner--Houghton's fixed
point equation in the local potential approximation, a Riccati equation, and
the Thomas--Fermi equation. We consider two models where the approach does
not apply in order to show the limitations of our Pad\'{e}--Hankel approach.
\end{abstract}

\section{Introduction\label{sec:Intro}}

Some time ago Fern\'{a}ndez et al \cite
{FMT89,F92,FG93,F95,F95b,F95c,F96,F96b,F97} developed a method for the
accurate calculation of eigenfunctions and eigenvalues for bound states and
resonances of the Schr\"{o}dinger equation. This approach is based on the
Taylor expansion of a regularized logarithmic derivative of the
eigenfunction. The physical eigenvalue is given by a sequence of roots of
Hankel determinants constructed from the coefficients of that series. One
merit of this approach, called Riccati--Pad\'{e}, is the great convergence
rate in most cases and that the same equation applies to bound states and
resonances. Besides, in some cases it yields upper and lower bounds to the
eigenvalues\cite{FMT89}.

The logarithmic derivative satisfies a Riccati equation, and one may wonder
if the method applies to other nonlinear ordinary differential equations.
The purpose of this paper is to investigate whether a kind of
Pad\'{e}--Hankel method may be useful for two--point boundary value problems
given by nonlinear ordinary differential equations.

In Section \ref{sec:Method} we outline the method, in Section \ref
{sec:Examples} we apply it to several problems of physical interest, and in
Section \ref{sec:Conclusions} we discuss the relative merits of the approach.

\section{Method\label{sec:Method}}

It is our purpose to propose a method for the treatment of two--point
boundary value problems. We suppose that the solution $f(x)$ of a nonlinear
ordinary differential equation can be expanded as
\begin{equation}
f(x)=x^{\alpha }\sum_{j=0}^{\infty }f_{j}x^{\beta j}  \label{eq:f_series}
\end{equation}
about $x=0$, where $\alpha $ and $\beta $ are real numbers, and $\beta >0$.
We also assume that we can calculate sufficient coefficients $f_{j}$ in
terms of one of them that should be determined by the boundary condition at
the other point; for example, at infinity. We show several illustrative
examples in the following section.

We try a rational approximation to $x^{-\alpha }f(x)$ of the form
\begin{equation}
\lbrack M,N](z)=\frac{\sum_{j=0}^{M}a_{j}z^{j}}{\sum_{j=0}^{N}b_{j}z^{j}}.
\label{eq:[M,N]}
\end{equation}
where $z=x^{\beta }$. The Taylor expansion of the usual Pad\'{e} approximant
yields $M+N+1$ coefficients of the series (\ref{eq:f_series})\cite{BO78};
but in the present case we require that the rational approximation (\ref
{eq:[M,N]}) gives us one more coefficient, that is to say, $M+N+2$. If $%
M=N+d $, $N=1,2,\ldots $, $d=0,1,\ldots $, this requirement leads to the
equation\cite{FMT89,F92,FG93,F95,F95b,F95c,F96,F96b,F97}
\begin{equation}
H_{D}^{d}=\left| f_{i+j+d+1}\right| _{i,j=0,1,\ldots N}=0,  \label{eq:Hankel}
\end{equation}
where $D=N+1=2,3,\ldots $ is the dimension of the Hankel determinant $%
H_{D}^{d}$.

In general, equation (\ref{eq:Hankel}) exhibits many roots and one expects
to find a sequence, for $D=2,3,\ldots $ and fixed $d$, that converges
towards the required value of the unknown coefficient. From now on we call
it Hankel sequence for short. If such convergent sequence is monotonously
increasing or decreasing we assume that it yields a lower or upper bound,
respectively. Such bounds where proved rigorously for some eigenvalue
problems\cite{FMT89}.

\section{Examples\label{sec:Examples}}

In order to test the performance of the Pad\'{e}--Hankel method, in this
section we consider the examples treated recently by Boisseau et al by means
of a most interesting algebraic approach\cite{BFG07}. We first consider the
field equation determining the vortex profile in a Ginzburg--Landau
effective theory\cite{BFG07} (and references therein)
\begin{equation}
f^{\prime \prime }(r)+\frac{1}{r}f^{\prime }(r)+\left( 1-\frac{n^{2}}{r^{2}}%
\right) f(r)-f(r)^{3}=0,\;r>0.  \label{eq:G-L}
\end{equation}
The solution $f(r)$ satisfies the expansion (\ref{eq:f_series}) with $x=r$, $%
\alpha =n=1,2,\ldots $, and $\beta =2$. If we substitute this series into
the differential equation and solve for the coefficients $f_{j}$, we obtain
them in terms of the only unknown $f_{0}$ that is determined by the boundary
condition at infinity: $f(r\rightarrow \infty )=1$\cite{BFG07} (and
references therein). The coefficients $f_{j}$, and therefore the Hankel
determinant $H_{D}^{d}$, are polynomial functions of $f_{0}$. For example,
for $n=1$ we have
\begin{equation}
f_{1}=-\frac{f_{0}}{8},\;f_{2}=\frac{f_{0}}{192}+\frac{f_{0}^{3}}{24}%
,\;f_{3}=-\frac{f_{0}}{9216}-\frac{5f_{0}^{3}}{576},\ldots
\end{equation}

Tables \ref{tab:GLn1} and \ref{tab:GLn2} show two Hankel sequences with $d=0$
and $d=1$ that converge rapidly towards the result of the accurate shooting
method\cite{BFG07} for $n=1$ and $n=2$, respectively. We appreciate that in
the case $n=1$ the sequences with $d=0$ and $d=1$ give upper and lower
bounds, respectively, that tightly bracket the exact value of the unknown
parameter of the theory: $0.58318949586060<f_{0}<0.58318949586061$.

On the other hand, the appropriate Hankel sequences are oscillatory when $%
n\geq 2$ and their rate of convergence decreases with $n$. Table \ref
{tab:GLn2-4} shows the best estimates of $f_{0}$ for $n=2,3,4$.

\begin{figure}[H]
\begin{center}
\includegraphics[width=9cm]{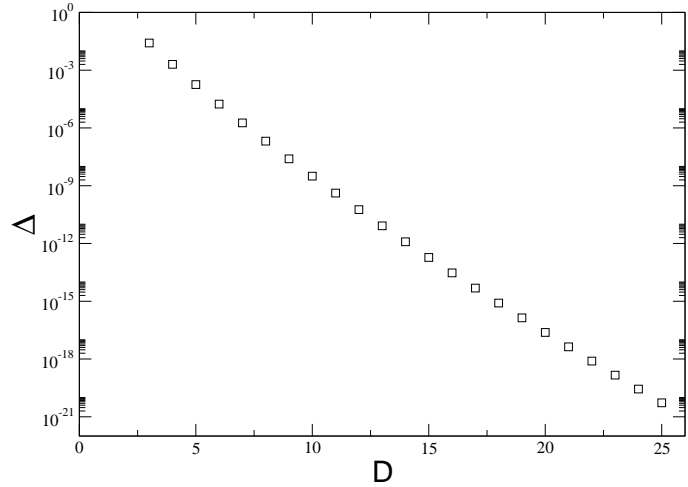}
\end{center}
\caption{$\Delta=\left|f_0(D,d=0)-f_0(D,d=1)\right|$ for Wilson's
renormalization}
\label{fig:Wilson}
\end{figure}

Our second example is the fixed--point equation for Wilson's exact
renormalization group\cite{BFG07} (and references therein)
\begin{equation}
2f^{\prime \prime }(x)-4f(x)f^{\prime }(x)-5xf^{\prime }(x)+f(x)=0,\;x>0.
\label{eq:W-RG}
\end{equation}
The solution to this equation can be expanded as in equation (\ref
{eq:f_series}) with $\alpha =1$ and $\beta =2$. The first coefficients are
\begin{equation}
f_{1}=\frac{f_{0}}{3}+\frac{f_{0}^{2}}{3},\;f_{2}=\frac{7f_{0}}{60}+\frac{%
f_{0}^{2}}{4}+\frac{2f_{0}^{3}}{15},\ldots .
\end{equation}
For large values of $x$ the physical solution should behave as $%
f(x)=ax^{1/5}+a^{2}/(5x^{3/5})+\ldots $. The Hankel sequences with $d=0$ and
$d=1$ converge towards the numerical result\cite{BFG07} (and references
therein) from above and below, respectively. Fig. \ref{fig:Wilson} displays
the great rate of convergence of these sequences as $\Delta =\left|
f_{0}(D,d=0)-f_{0}(D,d=1)\right| $, $D=2,3,\ldots $, from which we obtain
the accurate bounds $-1.22859820243702192438<f_{0}<-1.22859820243702192437$

The third example comes from a suitably modified Wegner--Houghton's fixed
point equation in the local potential approximation\cite{BFG07} (and
references therein)
\begin{equation}
2f^{\prime \prime }(x)+[1+f^{\prime }(x)][5f(x)-xf^{\prime }(x)]=0,\;x>0.
\label{eq:W-H}
\end{equation}
The solution satisfies the series (\ref{eq:f_series}) with $\alpha =1$ and $%
\beta =2$, and the first coefficients are
\begin{equation}
f_{1}=-\frac{f_{0}}{3}-\frac{f_{0}^{2}}{3},\;f_{2}=\frac{f_{0}}{60}+\frac{%
2f_{0}^{2}}{15}+\frac{7f_{0}^{3}}{60},\ldots .
\end{equation}
On the other hand, the acceptable solution should behave as $%
f(x)=ax^{5}-4/(3x)+\ldots $ when $x\gg 1$.

Table \ref{tab:WH} shows Hankel sequences with $d=0$ and $d=1$ that clearly
converge towards the numerical value of $f_{0}$\cite{BFG07} (and references
therein).

We have also applied our approach to the ordinary differential equation for
the spherically symmetric skyrmion field\cite{BFG07} (and references
therein) but we could not obtain convergent Hankel sequences. We do not know
yet the reason for the failure of the method in this case.

Present approach has earlier proved suitable for the treatment of the
Riccati equation derived from the Schr\"{o}dinger equation\cite
{FMT89,F92,FG93,F95,F95b,F95c,F96,F96b,F97}. Consider, for example, the
following Riccati equation
\begin{equation}
f^{\prime }(x)-f(x)^{2}+x^{2}=0,\;x>0.  \label{eq:Riccati}
\end{equation}
The solution can be expanded as in equation (\ref{eq:f_series}) with $\alpha
=\beta =1$; the first coefficients are
\[
f_{1}=f_{0}^{2},\;f_{2}=f_{0}^{3},\;f_{3}=-\frac{1}{3}+f_{0}^{4},\ldots
\]
There is a critical value $f_{0c}$ of $f(0)=f_{0}$ such that $f(x)\sim -x$
at large $x$ if $f(0)<f_{0c}$, $f(x)$ develops a singular point if $%
f(0)>f_{0c}$, and $f(x)\sim x$ at large $x$ if $f(0)=f_{0c}$. Present
Pad\'{e}--Hankel method yields the value of $f_{0c}$ with remarkable
accuracy as shown in Table \ref{tab:Riccati}. The rate of convergence of the
Hankel sequence for this problem is considerably greater than for the
preceding ones.

If we substitute $f(x)=-y^{\prime }(x)/y(x)$ into equation (\ref{eq:Riccati}%
), then the function $y(x)$ satisfies the Schr\"{o}dinger equation for a
harmonic oscillator with zero energy on the half line: $y^{\prime \prime
}(x)-x^{2}y(x)=0$, and the problem solved above is equivalent to finding the
logarithmic derivative at origin $y^{\prime }(0)/y(0)$ so that $y(x)$
behaves as $\exp (-x^{2}/2)$ at infinity. Obviously, any approach for linear
differential equations is suitable for this problem.

Finally, we consider the Thomas--Fermi equation \cite{BO78,PP87,FO90} (and
references therein)
\begin{equation}
\Phi ^{\prime \prime }(x)=x^{-1/2}\Phi (x)^{3/2},\;x>0  \label{eq:T-F}
\end{equation}
that provides a semiclassical description of the charge density in atoms of
high atomic number. It poses the problem of finding the slope at origin $%
\Phi ^{\prime }(0)$ so that $\Phi (0)=1$ and $\Phi (x\rightarrow \infty )=0$%
. The change of variables $t=x^{1/2}$ leads to a more tractable equation
\begin{equation}
t\xi ^{\prime \prime }(t)-\xi ^{\prime }(t)-4t^{2}\xi (t)^{3/2},\;t>0
\label{eq:T-F-xi}
\end{equation}
where $\xi (t)=\Phi (t^{2})$. We can expand the solution to this equation in
a Taylor series: $\xi (t)=1+a_{2}t^{2}+4t^{3}/3+2a_{2}t^{5}/5+\ldots $ and
the unknown slope is given by the unknown coefficient: $\Phi ^{\prime
}(0)=a_{2}$. One easily derives a recurrence relation for the coefficients
of this expansion\cite{FO90}. Although the treatment of equation (\ref
{eq:T-F-xi}) is straightforward, we find it more convenient to define the
function $f(t)=\xi (t)^{1/2}$ that is a solution to
\begin{equation}
t\left[ f(t)f^{\prime \prime }(t)+f^{\prime }(t)^{2}\right] -f(t)f^{\prime
}(t)-2t^{2}f(t)^{3}=0.  \label{eq:T-F-f}
\end{equation}
The function $f(t)$ satisfies a series like (\ref{eq:f_series}) with $\alpha
=0$ and $\beta =1$. The first coefficients are
\begin{equation}
f_{1}=0,\;f_{3}=\frac{2}{3},\;f_{4}=-\frac{f_{2}^{2}}{2},\;f_{5}=-\frac{%
4f_{2}}{15},\ldots
\end{equation}
We expect that a Hankel sequence will converge towards $f_{2}=f^{\prime
\prime }(0)/2=\Phi ^{\prime }(0)/2$. Since $f_{4}$ is the first nonzero
coefficient that depends on $f_{2}$ we choose $d\geq 4$ in the Hankel
determinant. Table \ref{tab:TF} shows that the Hankel sequence with $d=4$
converges rapidly giving a most accurate value of the slope at origin $\Phi
^{\prime }(0)$.

Finally, we consider two examples discussed by Bender et al\cite{BPW02}; the
first of them is the instanton equation
\begin{equation}
f^{\prime \prime }(x)+f(x)-f(x)^{3}=0  \label{eq:instanton}
\end{equation}
with the boundary conditions $f(0)=0$, $f(\infty )=1$. The solution to this
equation is $f(x)=\tanh \left( x/\sqrt{2}\right) $. The expansion of $f(x)$
is a particular case of equation (\ref{eq:f_series}) with $\alpha =1$ and $%
\beta =2$; its first coefficients being
\begin{equation}
f_{1}=-\frac{f_{0}}{6},\;\frac{f_{0}\left( 6f_{0}^{2}+1\right) }{120}%
,\;f_{3}=-\frac{f_{0}\left( 66f_{0}^{2}+1\right) }{5040},\ldots ,
\end{equation}
where $f_{0}=f^{\prime }(0)$ is the unknown. The Hankel series with $d=0$
and $d=1$ converge rapidly giving upper and lower bounds, respectively, to
the exact result $f_{0}=1/\sqrt{2}$.

The second example is the well known Blasius equation\cite{BPW02}
\begin{equation}
2y^{\prime \prime \prime }(x)+y(x)y^{\prime \prime }(x)=0  \label{eq:Blasius}
\end{equation}
with the boundary conditions $y(0)=y^{\prime }(0)=0$, $y^{\prime }(\infty )=1
$. The expansion of the solution in a Taylor series about $x=0$ is a
particular case of equation (\ref{eq:f_series}) with $\alpha =2$ and $\beta
=3$; its first coefficients are
\begin{equation}
f_{1}=-\frac{f_{0}^{2}}{60},\;f_{2}=\frac{11f_{0}^{3}}{20160},\ldots
\end{equation}
Since, in general,  $f_{j}\propto f_{0}^{j+1}$, then the only root of the
Hankel determinants is $f_{0}=0$ that leads to the trivial solution $%
y(x)\equiv 0$. We thus see another case where the Pad\'{e}--Hankel method
does not apply.

\section{Conclusions\label{sec:Conclusions}}

We have presented a simple method for the treatment of two--point boundary
value problems. If there is a suitable series for the solution about one
point, we construct a Hankel matrix with the expansion coefficients and
obtain the physical value of the undetermined coefficient from the roots of
a sequence of determinants. The value of this coefficient given by a
convergent Hankel sequence is exactly the one that produces the correct
asymptotic behaviour at the other point. We cannot prove this assumption
rigorously, but it seems that if there is a convergent sequence, it yields
the correct answer. Moreover, in some cases the Hankel sequences produce
upper and lower bounds bracketing the exact result tightly.

Present Pad\'{e}--Hankel approach is not as general as the one proposed by
Boisseau et al\cite{BFG07}, as we have already seen that the former does not
apparently apply to the skyrmion problem or to the Blasius equation\cite
{BPW02}. However, our procedure is much simpler and more straightforward and
may be a suitable alternative for the treatment of this kind of problems.
Besides, if our approach converges, it yields remarkably accurate results as
shown in the examples above. For example, it gives us the slope of the
electrostatic potential of the Thomas--Fermi theory with unprecedented
accuracy.

\begin{table}[H]
\caption{Convergence of the Hankel series for the
connection parameters of the global vortex for $n=1$}
\label{tab:GLn1}
\begin{tabular}{rll}
\hline
$D$ & $d=0$ & $d=1$ \\ \hline
2 & 0.595 & 0.578 \\
3 & 0.584 & 0.5829 \\
4 & 0.58324 & 0.58315 \\
5 & 0.58320 &  0.583183 \\
6 & 0.583192 & 0.583187 \\
7 & 0.583190 & 0.5831890 \\
8 & 0.5831897 &  0.5831893 \\
9 & 0.58318954 &  0.58318946 \\
10 & 0.58318952 & 0.58318948 \\
11 & 0.58318951 &  0.583189491 \\
12 & 0.583189498 &  0.583189494 \\
13 & 0.5831894964 & 0.5831894953 \\
14 & 0.5831894961 & 0.5831894956 \\
15 & 0.5831894960 &  0.5831894957 \\
16 & 0.58318949590 & 0.58318949583 \\
17 & 0.58318949588 & 0.58318949584 \\
18 & 0.583189495867 & 0.583189495854 \\
19 & 0.583189495864 & 0.583189495857 \\
20 & 0.583189495862 & 0.5831894958591 \\
21 & 0.5831894958609 & 0.5831894958598 \\
22 & 0.5831894958607 & 0.5831894958601
\end{tabular}
\end{table}

\begin{table}[H]
\caption{Convergence of the Hankel series for the
connection parameters of the global vortex for $n=2$}
\label{tab:GLn2}
\begin{tabular}{rll}
\hline
$D$ & $d=0$ & $d=1$ \\ \hline
3 & 0.156 &  0.151 \\
4 & 0.1528 & 0.154 \\
5 & 0.15310 & 0.1530 \\
6 & 0.15309 & 0.15311 \\
7 & 0.153098 & 0.15310 \\
8 & 0.1530997 & 0.15310 \\
9 & 0.1530991 & 0.153099 \\
10 & 0.15309914 & 0.1530989 \\
11 & 0.15309912 &  0.153099095 \\
12 & 0.15309917 &  0.153099091 \\
13 & 0.153099105 & 0.153099097 \\
14 & 0.1530991021 & 0.15309911 \\
15 & 0.15309910272 &  0.153099102 \\
16 & 0.153099102697 & 0.153099103 \\
17 & 0.153099102782 & 0.15309910292 \\
18 & 0.153099103124 & 0.15309910293 \\
19 & 0.153099102857 & 0.15309910289 \\
20 & 0.153099102864 & 0.15309910278 \\
21 & 0.15309910286136 & 0.153099102860 \\
22 & 0.15309910286142 & 0.153099102858
\end{tabular}
\end{table}

\begin{table}[H]
\caption{Best estimates of the connection parameters of the global vortex
for $n=2,3,4$ by means of Hankel sequences with $D\le D_{max}$}
\label{tab:GLn2-4}
\begin{center}
\begin{tabular}{lcl}
\hline
$n$ & $D_{max}$ & \multicolumn{1}{c}{$f_0$} \\ \hline
2 & 21 & 0.15309910286 \\
3 & 21 & 0.0261834207 \\
4 & 26 & 0.0033271734
\end{tabular}
\end{center}
\end{table}

\begin{table}[H]
\caption{Convergence of the Hankel sequences for the
Wegner--Houghton connection parameter}
\label{tab:WH}
\begin{center}
\begin{tabular}{cll}
\hline
$D$ & \multicolumn{1}{c}{$d=0$} & \multicolumn{1}{c}{$d=1$} \\ \hline
3 & -0.3013652092 & -0.4190129312 \\
4 & -0.5405112824 & -0.4696457170 \\
5 & -0.4552012493 & -0.4604796926 \\
6 & -0.4624525979 & -0.4616935821 \\
7 & -0.4613759926 & -0.4615091717 \\
8 & -0.4615571129 & -0.4615373393 \\
9 & -0.4615303767 & -0.4615331535 \\
10 & -0.4615342975 & -0.4615338165 \\
11 & -0.4615336147 & -0.4615337043 \\
12 & -0.4615337357 & -0.4615337227 \\
13 & -0.4615337173 & -0.4615337196 \\
14 & -0.4615337207 & -0.4615337202 \\
15 & -0.4615337200 & -0.4615337201 \\
16 & -0.46153372013 & -0.461533720119 \\
17 & -0.461533720113 & -0.4615337201157 \\
18 & -0.4615337201168 & -0.4615337201163 \\
19 & -0.4615337201161 & -0.4615337201162 \\
20 & -0.4615337201162 &
\end{tabular}
\end{center}
\end{table}

\begin{table}[H]
\caption{Convergence of the Hankel sequences with $d=0$ for the Riccati
equation.}
\label{tab:Riccati}
\begin{center}
\begin{tabular}{rl}
\hline
$D$ & \multicolumn{1}{c}{$f_0$} \\ \hline
4 & 0.6762 \\
5 & 0.675970 \\
6 & 0.6759785 \\
7 & 0.67597823 \\
8 & 0.6759782403 \\
9 & 0.675978240059 \\
10 & 0.6759782400675 \\
11 & 0.675978240067277 \\
12 & 0.6759782400672850 \\
13 & 0.675978240067284722 \\
14 & 0.675978240067284729 \\
15 & 0.67597824006728472899 \\
16 & 0.67597824006728472900 \\
17 & 0.67597824006728472900
\end{tabular}
\end{center}
\end{table}

\begin{table}[H]
\caption{Convergence of the Hankel sequences with $d=4$ for the
Thomas--Fermi equation}
\label{tab:TF}
\begin{center}
\begin{tabular}{rl}
\hline
$D$ & \multicolumn{1}{c}{$2f_2$} \\ \hline
10 & -1.5880709 \\
11 & -1.5880706 \\
12 & -1.58807103 \\
13 & -1.588071024 \\
14 & -1.5880710227 \\
15 & -1.58807102264 \\
16 & -1.588071022609 \\
17 & -1.588071022609 \\
18 & -1.5880710226116 \\
19 & -1.5880710226115 \\
20 & -1.58807102261139 \\
21 & -1.58807102261138 \\
22 & -1.58807102261137 \\
23 & -1.58807102261137 \\
24 & -1.5880710226113756 \\
25 & -1.58807102261137537 \\
26 & -1.58807102261137532 \\
27 & -1.5880710226113753154 \\
28 & -1.5880710226113753152 \\
29 & -1.5880710226113753154 \\
30 & -1.5880710226113753137
\end{tabular}
\end{center}
\end{table}

\end{document}